\documentclass[11pt,twoside]{article}
\usepackage{amsthm}
\usepackage{amsmath}
\usepackage{amssymb}
\usepackage{array}
\usepackage{graphicx}
\usepackage{graphics}
\usepackage{epsfig}
\usepackage{amsfonts,amscd}
\usepackage{exscale}
\usepackage{eucal}
\usepackage{latexsym,amsmath}
\usepackage{indentfirst}
\numberwithin{equation}{section}
\newtheorem{Prop}{\bf Proposition}[section]

\theoremstyle{definition} \theoremstyle{plain}

\begin{document}
\def \b{\Box}

\begin{center}
{\Large \bf Sample programs in $C^{++}$ for matrix computations in max plus algebra}\\[0.1cm]
\end{center}
\begin{center}
{\bf Mihai Ivan and Gheorghe Ivan} \\[0.1cm]
\end{center}

\setcounter{page}{1}

 \pagestyle{myheadings}

{\small {\bf Abstract}. The main purpose of this paper is to propose
five programs in $C^{++}$ for matrix computations and solving
recurrent equations systems with entries in max plus algebra.}
{\footnote{{\it AMS classification:} 15A80, 68-04.\\
{\it Key words and phrases:} idempotent semiring, max plus
algebra.}}

\section{Introduction}
\smallskip
\indent\indent Idempotent mathematics is based on replacing the
usual arithmetic operations with a new set of basic operations, that
is on replacing numerical fields by idempotent semirings. Exotic
semirings such as the max-plus algebra ${\bf R}_{max}$ have been
introduced in connection with various fields: graph theory, Markov
decision processes, discrete event systems theory, see \cite{kusa},
\cite{mohr}.

The paper is organized as follows. The semiring of matrices with
entries in max plus algebra is presented in Section 2.
 In Section 3 we give five programs in language $~C^{++}~$  for matrix computations in
 max plus algebra.

\section{Semirings. Matrices over max plus algebra}
\indent\indent We start this section by recalling of some necessary
backgrounds on semirings for our purposes (see
\cite{mohr},\cite{litv} and references therein for more details).

Let $S$ be a nonempty set endowed with two binary operations, {\it
addition} (denoted with $\oplus$) and {\it multiplication} (denoted
with $\otimes$). The algebraic structure $(S,\oplus, \otimes,
\varepsilon, e )$ is a \textit{semiring}, if it fulfills the
following conditions:

$(1)~~(S,\oplus, \varepsilon)~$ is a commutative monoid with
$\varepsilon$ as the neutral element for $\oplus;$

$(2)~(S, \otimes,  e )~$ is a  monoid with $\varepsilon$ as the
identity element for $\otimes;$

$(3)~~\otimes $ distributes over $\oplus;$

$(4)~~\varepsilon~$ is an absorbing element for $\otimes$, that is
$~a\otimes \varepsilon=\varepsilon \otimes a= \varepsilon,~\forall
a\in S.$

A semiring where addition is idempotent (that is, $~a\oplus
a=a,~\forall a \in S$) is called an {\it idempotent semiring}. If
$\otimes$ is commutative, we say that $S$ is a {\it commutative
semiring}.

Let $(S, \oplus, \otimes, \varepsilon, e )$ be an (idempotent)
semiring.  For each pair of positive integer $(m,n)$, let $
M_{m\times n}(S)$ be denote the set of $m\times n$ matrices with
entries in $S$. The operations $\oplus$ and $\otimes$ on $S$ induce
corresponding operations on $ M_{m\times n}(S)$ in the obvious way.
Indeed, if
$A=(A_{ij}), B=(B_{ij})\in M_{m\times n}(S)$ then we have:\\[-0.2cm]
\begin{equation}
A\oplus B= ((A\oplus B)_{ij})~~~\hbox{where}~~~ (A\oplus B)_{ij}:=
A_{ij} \oplus B_{ij}.\label{(2.1)}
\end{equation}
If $A=(A_{ij})\in M_{m\times n}(S)$ and $B=(B_{jk})\in M_{n\times p}(S)$ then we have:\\[-0.2cm]
\begin{equation}
A\otimes B= ((A\otimes B)_{ik}),~i =\overline{1,m},
~k=\overline{1,p}~~\hbox{where}~ (A\otimes B)_{ik}:=
\bigoplus\limits_{j=1}^{n} A_{ij}\otimes B_{jk}.\label{(2.2)}
\end{equation}

The product of a matrix $A=(A_{ij}) \in M_{m\times n}(S)$ with a
scalar
$\alpha\in S$ is given by:\\[-0.2cm]
\begin{equation}
\alpha \otimes A= ((\alpha\otimes A)_{ij})
~~~\hbox{where}~~~(\alpha\otimes A)_{ij}:= \alpha  \otimes
A_{ij}.\label{(2.3)}
\end{equation}

 The set $M_{n\times n}(S)$ contains two special matrices with entries in
$S$, namely the zero matrix $O_{\oplus n}$, which  has all its
entries equal to $\varepsilon$, and the identity matrix $I_{\otimes
n}$, which  has the diagonal entries equal to $e$ and the other
entries equal to $\varepsilon$.

It is easy to check that the following proposition holds.
\begin{Prop}
$( M_{n\times n}(S), \oplus, \otimes, O_{\oplus n}, I_{\otimes n}) $
is an idempotent semiring, where the operations $\oplus$ and
$\otimes$ are given in $(2.1)$ and $(2.2)$.\hfill$\Box$
\end{Prop}

\markboth{Mihai Ivan, Gheorghe Ivan}{Sample programs in $C^{++}$ for
matrix computations in max plus algebra}

We call $( M_{n\times n}(S), \oplus, \otimes, O_{\oplus n},
I_{\otimes n}) $ the {\it semiring of $n\times n$ matrices with
entries in $S$}. In particular, if $S:={\bf R}_{max}= ({\bf R}\cup
\{-\infty\},\oplus:=max, \otimes:=+, \varepsilon:=-\infty, e:= 0$ is
called the {\it semiring of $n\times n$ matrices over $ {\bf
R}_{max}$}.

When $S={\bf R}_{max}$, the operations $\oplus$ and
$\otimes$ given in $(2.1)$ and $(2.2)$, becomes:\\[-0.2cm]
\begin{equation}
(A\oplus B)_{ij}:= max\{A_{ij}, B_{ij}\}~~~\hbox{and}~~~ (A\otimes
B)_{ik}:= max_{1\leq k\leq n} \{A_{ij}+ B_{jk}\}.\label{(2.4)}
\end{equation}

The operation $\otimes$ on $ M_{m\times n}({\bf R}_{max})$
given in $(2.3)$ becomes:\\[-0.2cm]
\begin{equation}
\alpha \otimes A= ((\alpha\otimes A)_{ij})
~~~\hbox{where}~~~(\alpha\otimes A)_{ij}:= \alpha +
A_{ij}.\label{(2.5)}
\end{equation}

\section{Five programs in $C^{++}$}

In this section we give programs written in the language $C^{++}$
for the basic operations with matrices over ${\bf
R}_{max}$ and for solving a recurrent linear system:\\[0.2cm]
$\bf 1.~~~$ the sum of two matrices $A, B\in M_{m\times n}({\bf
R}_{max})$;\\[0.1cm]
$\bf 2.~~~$ the product of two matrices $A\in M_{m\times n}({\bf
R}_{max})$ and $B\in M_{n\times p}({\bf R}_{max})$;\\[0.1cm]
$\bf 3.~~~$ the product of a matrix $A\in M_{m\times n}({\bf
R}_{max})$
with a scalar $\alpha\in {\bf R}_{max}$;\\[0.1cm]
$\bf 4.~~~$ the power  of a matrix $A\in M_{n\times n}({\bf R}_{max})$;\\[0.1cm]
$\bf 5.~~~$ the solving a linear system of the form:
\[
X(k+1)=A\otimes X(k),~~~k\geq 0,
\]
where $~A\in M_{n\times n}({\bf R}_{max})$ and $X(k)\in M(n,1;{\bf
R}_{max}).$\\[0.1cm]

The principal program is constituted from the following
lines.\\[0.1cm]
using System;\\
using System.Collections.Generic;\\
using System.ComponentModel;\\
using System.Data;\\
using System.Drawing;\\
using System.Text;\\
using System.Windows.Forms;\\[0.1cm]
namespace Operations$_{-}$ with$_{-}$matrices\\
$\{$\\
public partial class Form1 : Form\\
$\{$\\
\hspace*{0.5cm}public Form1()\\
\hspace*{0.5cm}$\{$\\
\hspace*{1cm}InitializeComponent();\\
\hspace*{0.5cm}$\}$\\[0.1cm]
\hspace*{0.5cm}public int[,] A5 = new int[50, 50];\\[0.1cm]
$\#$region {\it sum of two matrices in max plus algebra}\\[0.1cm]
private void initMatrixA()\\
$\{$\\
\hspace*{0.5cm}int column = 0;\\
\hspace*{0.5cm}column = Convert.ToInt16(textColumn.Text);\\
\hspace*{0.5cm}dataGridA.ColumnCount = column;\\
\hspace*{0.5cm}dataGridA.AllowUserToOrderColumns = false;\\
\hspace*{0.5cm}dataGridA.AllowUserToAddRows = false;\\
\hspace*{0.5cm}dataGridA.Enabled = true;\\
\hspace*{0.5cm}dataGridA.AutoSizeRowsMode =
DataGridViewAutoSizeRowsMode.DisplayedCellsExceptHeaders;\\
\hspace*{0.5cm}dataGridA.ColumnHeadersBorderStyle =DataGridViewHeaderBorderStyle.Raised;\\
\hspace*{0.5cm}dataGridA.CellBorderStyle =DataGridViewCellBorderStyle.Single;\\
\hspace*{0.5cm}dataGridA.GridColor = Color.DodgerBlue;\\
\hspace*{0.5cm}dataGridA.ColumnHeadersVisible = false;\\
\hspace*{0.5cm}dataGridA.RowHeadersVisible = false;\\
\hspace*{0.5cm}dataGridA.BackgroundColor = Color.WhiteSmoke;\\
\hspace*{0.5cm}dataGridA.BorderStyle = BorderStyle.None;\\
\hspace*{0.5cm}dataGridA.AllowUserToResizeColumns = true;\\
\hspace*{0.5cm}DataGridViewCellStyle style = new DataGridViewCellStyle();\\
\hspace*{0.5cm}style.Format = "N0";\\
\hspace*{0.5cm}style.NullValue = null;\\[0.1cm]
\hspace*{0.5cm}DataGridViewCellStyle columnHeaderStyle = new DataGridViewCellStyle();\\
\hspace*{0.5cm}columnHeaderStyle.BackColor = Color.AntiqueWhite;\\
\hspace*{0.5cm}columnHeaderStyle.Alignment = DataGridViewContentAlignment.MiddleCenter;\\
\hspace*{0.5cm}dataGridA.ColumnHeadersDefaultCellStyle =
            columnHeaderStyle;\\[0.1cm]
\hspace*{0.5cm}DataGridViewCell CellR = new DataGridViewTextBoxCell();\\
\hspace*{0.5cm}CellR.Style.Alignment =
            DataGridViewContentAlignment.MiddleRight;\\[0.2cm]
\hspace*{0.5cm}DataGridViewCell CellL = new DataGridViewTextBoxCell();\\
\hspace*{0.5cm}CellL.Style.Alignment =
            DataGridViewContentAlignment.MiddleLeft;\\[0.1cm]
\hspace*{0.5cm}int line = 0;\\
\hspace*{0.5cm}line = Convert.ToInt16(textLine.Text);\\
\hspace*{0.5cm}dataGridA.RowCount = line;\\[0.1cm]
\hspace*{0.5cm}for (int i = 0; i $<$ column; i++)\\
\hspace*{0.5cm}$\{$\\
\hspace*{1cm}dataGridA.Columns[i].Name = "C" + (i + 1);\\
\hspace*{1cm}dataGridA.Columns[i].CellTemplate = CellR;\\
\hspace*{1cm}dataGridA.Columns[i].Width = 30;\\
\hspace*{1cm}dataGridA.Columns[i].DefaultCellStyle = style;\\
\hspace*{0.5cm}$\}$\\
$\}$\\[0.1cm]
private void initMatrixB()\\
$\{$\\
\hspace*{0.5cm}int column = 0;\\
\hspace*{0.5cm} column = Convert.ToInt16(textColumn.Text);\\
\hspace*{0.5cm}dataGridB.ColumnCount = column;\\
\hspace*{0.5cm}dataGridB.AllowUserToOrderColumns = false;\\
\hspace*{0.5cm}dataGridB.AllowUserToAddRows = false;\\
\hspace*{0.5cm}dataGridB.Enabled = true;\\
\hspace*{0.5cm}dataGridB.AutoSizeRowsMode = DataGridViewAutoSizeRowsMode.DisplayedCellsExceptHeaders;\\
\hspace*{0.5cm}dataGridB.ColumnHeadersBorderStyle = DataGridViewHeaderBorderStyle.Raised;\\
\hspace*{0.5cm}dataGridB.CellBorderStyle = DataGridViewCellBorderStyle.Single;\\
\hspace*{0.5cm}dataGridB.GridColor = Color.DodgerBlue;\\
\hspace*{0.5cm}dataGridB.ColumnHeadersVisible = false;\\
\hspace*{0.5cm}dataGridB.RowHeadersVisible = false;\\
\hspace*{0.5cm}dataGridB.BackgroundColor = Color.WhiteSmoke;\\
\hspace*{0.5cm}dataGridB.BorderStyle = BorderStyle.None;\\
            \hspace*{0.5cm}dataGridB.AllowUserToResizeColumns = true;\\[0.2cm]
\hspace*{0.5cm}DataGridViewCellStyle columnHeaderStyle = new DataGridViewCellStyle();\\
\hspace*{0.5cm}columnHeaderStyle.BackColor = Color.AntiqueWhite;\\
\hspace*{0.5cm}columnHeaderStyle.Alignment = DataGridViewContentAlignment.MiddleCenter;\\
\hspace*{0.5cm}dataGridB.ColumnHeadersDefaultCellStyle = columnHeaderStyle;\\[0.2cm]
\hspace*{0.5cm}DataGridViewCell CellR = new DataGridViewTextBoxCell();\\
\hspace*{0.5cm}CellR.Style.Alignment = DataGridViewContentAlignment.MiddleRight;\\[0.1cm]
\hspace*{0.5cm}DataGridViewCell CellL = new DataGridViewTextBoxCell();\\
\hspace*{0.5cm}CellL.Style.Alignment = DataGridViewContentAlignment.MiddleLeft;\\[0.1cm]
\hspace*{0.5cm}int line = 0;\\
\hspace*{0.5cm}line = Convert.ToInt16(textLine.Text);\\
\hspace*{0.5cm}dataGridB.RowCount = line;\\[0.1cm]
\hspace*{0.5cm}for (int i = 0; i $<$ column; i++)\\
\hspace*{0.5cm}$\{$\\
\hspace*{1cm}dataGridB.Columns[i].Name = "C" + (i + 1);\\
\hspace*{1cm}dataGridB.Columns[i].CellTemplate = CellR;\\
\hspace*{0.5cm}dataGridB.Columns[i].Width = 30;\\
\hspace*{0.5cm}$\}$\\
$\}$\\[0.1cm]
private void initMatrix of Addition Results ()\\
$\{$\\
\hspace*{0.5cm}int column = 0;\\
\hspace*{0.5cm}column = Convert.ToInt16(textColumn.Text);\\
\hspace*{0.5cm}dataGridResAddition.ColumnCount = column;\\
\hspace*{0.5cm}dataGridResAddition.AllowUserToOrderColumns = false;\\
\hspace*{0.5cm}dataGridResAddition.AllowUserToAddRows = false;\\
\hspace*{0.5cm}dataGridResAddition.Enabled = true;\\
\hspace*{0.5cm}dataGridResAddition.AutoSizeRowsMode = DataGridViewAutoSizeRowsMode.DisplayedCellsExceptHeaders;\\
\hspace*{0.5cm}dataGridResAddition.ColumnHeadersBorderStyle = DataGridViewHeaderBorderStyle.Raised;\\
\hspace*{0.5cm}dataGridResAddition.CellBorderStyle = DataGridViewCellBorderStyle.Single;\\
\hspace*{0.5cm}dataGridResAddition.GridColor = Color.DodgerBlue;\\
\hspace*{0.5cm}dataGridResAddition.ColumnHeadersVisible = false;\\
\hspace*{0.5cm}dataGridResAddition.RowHeadersVisible = false;\\
\hspace*{0.5cm}dataGridResAddition.BackgroundColor = Color.WhiteSmoke;\\
\hspace*{0.5cm}dataGridResAddition.BorderStyle = BorderStyle.None;\\
\hspace*{0.5cm}dataGridResAddition.AllowUserToResizeColumns = true;\\[0.2cm]
\hspace*{0.5cm}DataGridViewCellStyle columnHeaderStyle = new DataGridViewCellStyle();\\
\hspace*{0.5cm} columnHeaderStyle.BackColor = Color.AntiqueWhite;\\
\hspace*{0.5cm}columnHeaderStyle.Alignment = DataGridViewContentAlignment.MiddleCenter;\\
\hspace*{0.5cm}dataGridResAddition.ColumnHeadersDefaultCellStyle = columnHeaderStyle;\\[0.1cm]
\hspace*{0.5cm}DataGridViewCell CellR = new DataGridViewTextBoxCell();\\
\hspace*{0.5cm}CellR.Style.Alignment = DataGridViewContentAlignment.MiddleRight;\\[0.1cm]
 \hspace*{0.5cm}DataGridViewCell CellL = new DataGridViewTextBoxCell();\\
 \hspace*{0.5cm}CellL.Style.Alignment = DataGridViewContentAlignment.MiddleLeft;\\[0.1cm]
 \hspace*{0.5cm}int line = 0;\\
 \hspace*{0.5cm}line = Convert.ToInt16(textLine.Text);\\
 \hspace*{0.5cm}dataGridResAddition.RowCount = line;\\
 \hspace*{0.5cm}for (int i = 0; i $<$ column; i++)\\
 \hspace*{0.5cm}$\{$\\
 \hspace*{1cm}dataGridResAddition.Columns[i].Name = "C" + (i + 1);\\
 \hspace*{1cm}dataGridResAddition.Columns[i].CellTemplate = CellR;\\
 \hspace*{1cm}dataGridResAddition.Columns[i].Width = 30;\\
 \hspace*{0.5cm}$\}$\\
 $\}$\\
 private void btGenerate$_{-}$Click(object sender, EventArgs e)\\
 $\{$\\
 \hspace*{0.5cm}if (textLine.Text != "" $\&\&$ textColumn.Text != "")\\
 \hspace*{0.5cm}$\{$\\
 \hspace*{1cm}initMatrixA();\\
 \hspace*{1cm}initMatrixB();\\
 \hspace*{0.5cm}$\}$\\
 \hspace*{0.5cm}else\\
 \hspace*{0.5cm}$\{$\\
 \hspace*{1cm}if (textLine.Text == "" $\&\&$ textColumn.Text == "")\\
 \hspace*{1.5cm}MessageBox.Show("Introduce number of lines and number of columns!");\\
 \hspace*{1cm}else if (textLine.Text == "")\\
 \hspace*{1.5cm}MessageBox.Show("Introduce number of lines!");\\
 \hspace*{1cm}else if (textColumns.Text == "")\\
 \hspace*{1.5cm}MessageBox.Show("Introduce number of columns!");\\
 \hspace*{0.5cm}$\}$\\
 $\}$\\
 private void btComputeSumClick(object sender, EventArgs e)\\
 $\{$\\
 \hspace*{0.5cm}initMatrixAdditionResult();\\[0.1cm]
 \hspace*{0.5cm}int line = 0;\\
 \hspace*{0.5cm}int column = 0;\\
 \hspace*{0.5cm}column = Convert.ToInt16(textColumn.Text);\\
 \hspace*{0.5cm}line = Convert.ToInt16(textLine.Text);\\
 \hspace*{0.5cm}int[,] A = new int[line, column];\\
 \hspace*{0.5cm}int[,] B = new int[line, column];\\
 \hspace*{0.5cm}int[,] ResAddition = new int[line, column];\\
 \hspace*{0.5cm}$\#$ region reading matrix A\\
 \hspace*{0.5cm}for (int i = 0; i $<$ line; i++)\\
 \hspace*{0.5cm}$\{$\\
 \hspace*{1cm}for (int j = 0; j $<$ column; j++)\\
 \hspace*{1cm}$\{$\\
 \hspace*{1.5cm}if (dataGridA.Rows[i].Cells[j].Value.ToString() == "E")\\
 \hspace*{1.8cm} A[i, j] = Int32.MinValue;\\
 \hspace*{1.5cm}else\\
 \hspace*{1.8cm} A[i, j] = Convert.ToInt16(dataGridA.Rows[i].Cells[j].Value.ToString());\\
 \hspace*{1.5cm}if (dataGridB.Rows[i].Cells[j].Value.ToString() == "E")\\
 \hspace*{1.8cm}B[i, j] = Int32.MinValue;\\
 \hspace*{1.5cm}else\\
 \hspace*{1.8cm}B[i, j] = Convert.ToInt16(dataGridB.Rows[i].Cells[j].Value.ToString());\\
 \hspace*{1cm}$\}$\\
 \hspace*{0.5cm}$\}$\\
 \hspace*{0.5cm}$\#$ endregion\\
 \hspace*{0.5cm}$\#$ region computation of sum\\
 \hspace*{0.5cm}for (int i = 0; i $<$ line; i++)\\
 \hspace*{0.5cm}$\{$\\
 \hspace*{1cm}for (int j = 0; j $<$ column; j++)\\
 \hspace*{1cm}$\{$\\
 \hspace*{1.5cm} if (A[i, j] $>$ B[i, j])\\
 \hspace*{1.8cm}ResAddition[i, j] = A[i, j];\\
 \hspace*{1.5cm}else\\
 \hspace*{1.8cm}ResAddition[i, j] = B[i, j];\\
 \hspace*{1cm}$\}$\\
 \hspace*{0.5cm}$\}$\\
 $\#$ endregion\\
 \hspace*{0.5cm}$\#$ region display sum\\
 \hspace*{0.5cm}for (int i = 0; i $<$ line; i++)\\
 \hspace*{0.5cm}$\{$\\
 \hspace*{1cm}for (int j = 0; j $<$ column; j++)\\
 \hspace*{1cm}$\{$\\
 \hspace*{1.5cm}if (ResAddition[i, j] == Int32.MinValue)\\
 \hspace*{1.8cm}dataGridResAddition.Rows[i].Cells[j].Value = "E";\\
 \hspace*{1.5cm}else\\
 \hspace*{1.8cm}  dataGridResAddition.Rows[i].Cells[j].Value = ResAddition[i, j];\\
 \hspace*{1cm}$\}$\\
 \hspace*{0.5cm}$\}$\\
 \hspace*{0.5cm}$\#$endregion\\
 $\}$\\
 public void Reset$_{-}$ Values $_{-}$ For $_{-}$ Addition()\\
 $\{$\\
 \hspace*{0.5cm}textColumn.ResetText();\\
 \hspace*{0.5cm}textLine.ResetText();\\
 \hspace*{0.5cm}dataGridA.Rows.Clear();\\
 \hspace*{0.5cm}dataGridB.Rows.Clear();\\
 \hspace*{0.5cm}dataGridResAddition.Rows.Clear();\\
 $\}$\\
 private void btReset$_{-}$Click(object sender, EventArgs e)\\
 $\{$\\
 \hspace*{0.5cm}Reset$_{-}$Values $_{-}$For$_{-}$Addition();\\
 $\}$\\
$\#$ endregion\\[0.3cm]
$\#$ region {\it product of two matrices in max plus
algebra}\\[0.1cm]
private void initMatrixA2()\\
$\{$\\
\hspace*{0.5cm}int column = 0;\\
\hspace*{0.5cm}column = Convert.ToInt16(textColumnA2.Text);\\
\hspace*{0.5cm}dataGridA2.ColumnCount = column;\\
\hspace*{0.5cm}dataGridA2.AllowUserToOrderColumns = false;\\
\hspace*{0.5cm}dataGridA2.AllowUserToAddRows = false;\\
\hspace*{0.5cm}dataGridA2.Enabled = true;\\
\hspace*{0.5cm}dataGridA2.AutoSizeRowsMode =\\
 DataGridViewAutoSizeRowsMode.DisplayedCellsExceptHeaders;\\
\hspace*{0.5cm}dataGridA2.ColumnHeadersBorderStyle = DataGridViewHeaderBorderStyle.Raised;\\
\hspace*{0.5cm}dataGridA2.CellBorderStyle = DataGridViewCellBorderStyle.Single;\\
\hspace*{0.5cm}dataGridA2.GridColor = Color.DodgerBlue;\\
\hspace*{0.5cm}dataGridA2.ColumnHeadersVisible = false;\\
\hspace*{0.5cm}dataGridA2.RowHeadersVisible = false;\\
\hspace*{0.5cm}dataGridA2.BackgroundColor =Color.WhiteSmoke;\\
\hspace*{0.5cm}dataGridA2.BorderStyle = BorderStyle.None;\\
\hspace*{0.5cm}dataGridA2.AllowUserToResizeColumns = true;\\[0.1cm]
\hspace*{0.5cm}DataGridViewCellStyle columnHeaderStyle = new DataGridViewCellStyle();\\
\hspace*{0.5cm}columnHeaderStyle.BackColor = Color.AntiqueWhite;\\
\hspace*{0.5cm}columnHeaderStyle.Alignment = DataGridViewContentAlignment.MiddleCenter;\\
\hspace*{0.5cm}dataGridA2.ColumnHeadersDefaultCellStyle = columnHeaderStyle;\\[0.1cm]
\hspace*{0.5cm}DataGridViewCell CellR = new DataGridViewTextBoxCell();\\
\hspace*{0.5cm}CellR.Style.Alignment = DataGridViewContentAlignment.MiddleRight;\\[0.1cm]
\hspace*{0.5cm}DataGridViewCell CellL = new DataGridViewTextBoxCell();\\
\hspace*{0.5cm}CellL.Style.Alignment = DataGridViewContentAlignment.MiddleLeft;\\[0.1cm]
\hspace*{0.5cm}int line = 0;\\
\hspace*{0.5cm}line = Convert.ToInt16(textLineA2.Text);\\
\hspace*{0.5cm}dataGridA2.RowCount = line;\\[0.1cm]
\hspace*{0.5cm}for (int i = 0; i $<$ column; i++)\\
\hspace*{0.5cm}$\{$\\
\hspace*{1cm}dataGridA2.Columns[i].CellTemplate = CellR;\\
\hspace*{1cm}dataGridA2.Columns[i].Width = 30;\\
\hspace*{0.5cm}$\}$\\
$\}$\\
private void initMatrixB2()\\
$\{$\\
\hspace*{0.5cm}int column = 0;\\
\hspace*{0.5cm}column = Convert.ToInt16(textcolumnB2.Text);\\
\hspace*{0.5cm}dataGridB2.ColumnCount = column;\\
\hspace*{0.5cm}dataGridB2.AllowUserToOrderColumns = false;\\
\hspace*{0.5cm}dataGridB2.AllowUserToAddRows = false;\\
\hspace*{0.5cm}dataGridB2.Enabled = true;\\
\hspace*{0.5cm}dataGridB2.AutoSizeRowsMode =\\
 DataGridViewAutoSizeRowsMode.DisplayedCellsExceptHeaders;\\
\hspace*{0.5cm}dataGridB2.ColumnHeadersBorderStyle = DataGridViewHeaderBorderStyle.Raised;\\
\hspace*{0.5cm}dataGridB2.CellBorderStyle = DataGridViewCellBorderStyle.Single;\\
\hspace*{0.5cm}dataGridB2.GridColor = Color.DodgerBlue;\\
\hspace*{0.5cm}dataGridB2.ColumnHeadersVisible = false;\\
\hspace*{0.5cm}dataGridB2.RowHeadersVisible = false;\\
\hspace*{0.5cm}dataGridB2.BackgroundColor = Color.WhiteSmoke;\\
\hspace*{0.5cm}dataGridB2.BorderStyle = BorderStyle.None;\\
\hspace*{0.5cm}dataGridB2.AllowUserToResizeColumns = true;\\[0.1cm]
\hspace*{0.5cm}DataGridViewCellStyle columnHeaderStyle = new DataGridViewCellStyle();\\
\hspace*{0.5cm}columnHeaderStyle.BackColor = Color.AntiqueWhite;\\
\hspace*{0.5cm}columnHeaderStyle.Alignment = DataGridViewContentAlignment.MiddleCenter;\\
\hspace*{0.5cm}dataGridB2.ColumnHeadersDefaultCellStyle = columnHeaderStyle;\\[0.1cm]
\hspace*{0.5cm}DataGridViewCell CellR = new DataGridViewTextBoxCell();\\
\hspace*{0.5cm}CellR.Style.Alignment = DataGridViewContentAlignment.MiddleRight;\\[0.1cm]
\hspace*{0.5cm}DataGridViewCell CellL = new DataGridViewTextBoxCell();\\
\hspace*{0.5cm}CellL.Style.Alignment = DataGridViewContentAlignment.MiddleLeft;\\[0.1cm]
\hspace*{0.5cm}int line = 0;\\
\hspace*{0.5cm}line = Convert.ToInt16(textLineB2.Text);\\
\hspace*{0.5cm}dataGridB2.RowCount = line;\\[0.1cm]
\hspace*{0.5cm}for (int i = 0; i $<$ column; i++)\\
\hspace*{0.5cm}$\{$\\
\hspace*{1cm}dataGridB2.Columns[i].Name = "C" + (i + 1);\\
\hspace*{1cm}dataGridB2.Columns[i].CellTemplate = CellR;\\
\hspace*{1cm}dataGridB2.Columns[i].Width = 30;\\
\hspace*{0.5cm}$\}$\\
$\}$\\
private void initMatrixProduct()\\
$\{$\\
\hspace*{0.5cm}int column = 0;\\
\hspace*{0.5cm}column = Convert.ToInt16(textcolumnB2.Text);\\
\hspace*{0.5cm}dataGridProduct.ColumnCount = column;\\
\hspace*{0.5cm}dataGridProduct.AllowUserToOrderColumns = false;\\
\hspace*{0.5cm}dataGridProduct.AllowUserToAddRows = false;\\
\hspace*{0.5cm}dataGridProduct.Enabled = true;\\
\hspace*{0.5cm}dataGridProduct.AutoSizeRowsMode =\\
 DataGridViewAutoSizeRowsMode.DisplayedCellsExceptHeaders;\\
\hspace*{0.5cm}dataGridProduct.ColumnHeadersBorderStyle = DataGridViewHeaderBorderStyle.Raised;\\
\hspace*{0.5cm}dataGridProduct.CellBorderStyle = DataGridViewCellBorderStyle.Single;\\
\hspace*{0.5cm}dataGridProduct.GridColor = Color.DodgerBlue;\\
\hspace*{0.5cm}dataGridProduct.ColumnHeadersVisible = false;\\
\hspace*{0.5cm}dataGridProduct.RowHeadersVisible = false;\\
\hspace*{0.5cm}dataGridProduct.BackgroundColor = Color.WhiteSmoke;\\
\hspace*{0.5cm}dataGridProduct.BorderStyle = BorderStyle.None;\\
\hspace*{0.5cm}dataGridProduct.AllowUserToResizeColumns = true;\\[0.1cm]
\hspace*{0.5cm}DataGridViewCellStyle columnHeaderStyle = new DataGridViewCellStyle();\\
\hspace*{0.5cm}columnHeaderStyle.BackColor = Color.AntiqueWhite;\\
\hspace*{0.5cm} columnHeaderStyle.Alignment = DataGridViewContentAlignment.MiddleCenter;\\
\hspace*{0.5cm}dataGridProduct.ColumnHeadersDefaultCellStyle = columnHeaderStyle;\\[0.1cm]
\hspace*{0.5cm}DataGridViewCell CellR = new DataGridViewTextBoxCell();\\
\hspace*{0.5cm}CellR.Style.Alignment = DataGridViewContentAlignment.MiddleRight;\\[0.1cm]
\hspace*{0.5cm}DataGridViewCell CellL = new DataGridViewTextBoxCell();\\
\hspace*{0.5cm}CellL.Style.Alignment = DataGridViewContentAlignment.MiddleLeft;\\[0.1cm]
\hspace*{0.5cm}int line = 0;\\
\hspace*{0.5cm}line = Convert.ToInt16(textLineA2.Text);\\
\hspace*{0.5cm}dataGridProduct.RowCount = line;\\[0.1cm]
\hspace*{0.5cm}for (int i = 0; i $<$ column; i++)\\
\hspace*{0.5cm}$\{$\\
\hspace*{1cm}dataGridProduct.Columns[i].Name = "C" + (i + 1);\\
\hspace*{1cm}dataGridProduct.Columns[i].CellTemplate = CellR;\\
\hspace*{1cm}dataGridProduct.Columns[i].Width = 30;\\
\hspace*{0.5cm}$\}$\\
$\}$\\
private void textColumnA2$_{-}$Leave(object sender, EventArgs e)\\
$\{$\\
\hspace*{0.5cm}textLineB2.Text = textColumnA2.Text;\\
$\}$\\
private void btGenerare2$_{-}$Click(object sender, EventArgs e)\\
$\{$\\
\hspace*{0.5cm}if (textLineA2.Text != "" $\&\&$ textColumnA2.Text != "" $\&\&$ textLineB2.Text != "" $\&\&$ textcolumnB2.Text != "")\\
\hspace*{0.5cm}$\{$\\
\hspace*{1cm}initMatrixA2();\\
\hspace*{1cm}initMatrixB2();\\
\hspace*{0.5cm}$\}$\\
\hspace*{0.5cm}else\\
\hspace*{0.5cm}$\{$\\
\hspace*{1cm} if (textLineA2.Text == "" $\&\&$ textColumnA2.Text == "" $\&\&$ textLineB2.Text == "" $\&\&$ textcolumnoB2.Text == "")\\
\hspace*{1.5cm} MessageBox.Show("Introduce number of lines and number of columns!");\\
\hspace*{1cm}else\\
\hspace*{1cm}$\{$\\
\hspace*{1.5cm}if (textLineA2.Text == "" $||$ textLineB2.Text == "")\\
\hspace*{1.8cm}MessageBox.Show("Introduce number of lines for matrix A!");\\
\hspace*{1.5cm}else if (textColumnA2.Text == "")\\
\hspace*{1.8cm}MessageBox.Show("Introduce number of columns for matrix A!");\\
\hspace*{1cm}else if (textcolumnoB2.Text == "")\\
\hspace*{1.5cm}MessageBox.Show("Introduce number of columns for matrix B!");\\
\hspace*{1cm}$\}$\\
\hspace*{0.5cm}$\}$\\
$\}$\\
private void btComputationProduct$_{-}$Click(object sender, EventArgs e)\\
$\{$\\
\hspace*{0.5cm}initMatrixProduct();\\[0.1cm]
\hspace*{0.5cm}int lineA = 0; int lineB = 0;\\
\hspace*{0.5cm}int columnA = 0; int columnB = 0;\\
\hspace*{0.5cm}columnA = Convert.ToInt16(textColumnA2.Text);\\
\hspace*{0.5cm}lineA = Convert.ToInt16(textLineA2.Text);\\
\hspace*{0.5cm}columnB = Convert.ToInt16(textcolumnB2.Text);\\
\hspace*{0.5cm}lineB = Convert.ToInt16(textLineB2.Text);\\
\hspace*{0.5cm}int[,] A2 = new int[lineA, columnA];\\
\hspace*{0.5cm}int[,] B2 = new int[lineB, columnB];\\
\hspace*{0.5cm}int[,] Product = new int[lineA, columnB];\\
\hspace*{0.5cm}int[,] Sum = new int[lineA, columnaB];\\
\hspace*{0.5cm}int k;\\
\hspace*{0.5cm}$\#$ region reading matrices A and B\\
\hspace*{0.5cm}for (int i = 0; i $<$ lineA; i++)\\
\hspace*{0.5cm}$\{$\\
\hspace*{1cm}for (int j = 0; j $<$ columnA; j++)\\
\hspace*{1cm}$\{$\\
\hspace*{1cm}if (dataGridA2.Rows[i].Cells[j].Value.ToString() == "E")\\
\hspace*{1.5cm}A2[i, j] = Int32.MinValue;\\
\hspace*{1cm}else\\
\hspace*{1.5cm}A2[i, j] = Convert.ToInt16(dataGridA2.Rows[i].Cells[j].Value.ToString());\\
\hspace*{1cm}$\}$\\
\hspace*{0.5cm}$\}$\\[0.1cm]
\hspace*{0.5cm}for (int i = 0; i $<$ lineB; i++)\\
\hspace*{0.5cm}$\{$\\
\hspace*{1cm}for (int j = 0; j $<$ columnB; j++)\\
\hspace*{1cm}$\{$\\
\hspace*{1cm}if (dataGridB2.Rows[i].Cells[j].Value.ToString() == "E")\\
\hspace*{1.5cm}B2[i, j] = Int32.MinValue;\\
\hspace*{1cm}else\\
\hspace*{1.5cm}B2[i, j] = Convert.ToInt16(dataGridB2.Rows[i].Cells[j].Value.ToString());\\
\hspace*{1cm}$\}$\\
\hspace*{0.5cm}$\}$\\
\hspace*{0.5cm}$\#$ endregion\\
\hspace*{0.5cm}$\#$ region computation product\\
\hspace*{0.5cm}for (int i = 0; i $<$ lineA; i++)
\hspace*{0.5cm}$\{$\\
\hspace*{1cm}for (int j = 0; j $<$ columnB; j++)\\
\hspace*{1cm}$\{$\\
\hspace*{1.5cm}Product[i, j] = Int32.MinValue;\\
\hspace*{1.5cm}for (k = 0; k $<$ lineB; k++)\\
\hspace*{1.5cm}$\{$\\
\hspace*{1.8cm} if (A2[i, k] == Int32.MinValue $||$ B2[k, j] == Int32.MinValue)\\
\hspace*{1.8cm}Sum[i, j] = Int32.MinValue;\\
\hspace*{1.5cm} else\\
\hspace*{1.8cm}Sum[i, j] = A2[i, k] + B2[k, j];\\[0.1cm]
\hspace*{1.5cm}if (Product[i, j] $>$ Sum[i, j])\\
\hspace*{1.8cm}Product[i, j] = Product[i, j];\\
\hspace*{1.5cm}else\\
\hspace*{1.8cm}Product[i, j] = Sum[i, j];\\
\hspace*{1.5cm}$\}$\\
\hspace*{1cm}$\}$\\
\hspace*{0.5cm}$\}$\\
\hspace*{0.5cm}$\#$endregion\\
\hspace*{0.5cm}$\#$region display product\\
\hspace*{0.5cm}for (int i = 0; i $<$ lineA; i++)\\
\hspace*{0.5cm}$\{$\\
\hspace*{1cm}for (int j = 0; j $<$ columnB; j++)\\
\hspace*{1cm}$\{$\\
\hspace*{1.5cm} if (Product[i, j] == Int32.MinValue)\\
\hspace*{1.8cm} dataGridProduct.Rows[i].Cells[j].Value = "E";\\
\hspace*{1.5cm}else\\
\hspace*{1.8cm}dataGridProduct.Rows[i].Cells[j].Value = Product[i, j];\\
\hspace*{1.5cm}$\}$\\
\hspace*{0.5cm}$\}$\\
\hspace*{0.5cm}$\#$ endregion\\
$\}$\\
public void Reset$_{-}$Values for Multiplication()\\
$\{$\\
\hspace*{0.5cm}textColumnA2.ResetText();\\
\hspace*{0.5cm}textLineA2.ResetText();\\
\hspace*{0.5cm}textcolumnB2.ResetText();\\
\hspace*{0.5cm}textLineB2.ResetText();\\
\hspace*{0.5cm}dataGridA2.Rows.Clear();\\
\hspace*{0.5cm}dataGridB2.Rows.Clear();\\
\hspace*{0.5cm}dataGridProduct.Rows.Clear();\\
$\}$\\
private void btResetMultiplication$_{-}$Click(object sender, EventArgs e)\\
$\{$\\
\hspace*{0.5cm}Reset$_{-}$Values$_{-}$for$_{-}$Multiplication ();\\
$\}$\\
$\#$ endregion\\[0.2cm]
$\#$region {\it scalar product in max plus algebra}\\[0.1cm]
private void initMatrixA3()\\
$\{$\\
\hspace*{0.5cm} int column = 0;\\
\hspace*{0.5cm}column =
            Convert.ToInt16(textcolumnA3.Text);\\
\hspace*{0.5cm}dataGridA3.ColumnCount = column;\\
\hspace*{0.5cm}dataGridA3.AllowUserToOrderColumns =
            false;\\
\hspace*{0.5cm}dataGridA3.AllowUserToAddRows = false;\\
\hspace*{0.5cm}dataGridA3.Enabled = true;\\
\hspace*{0.5cm}dataGridA3.AutoSizeRowsMode =\\
            DataGridViewAutoSizeRowsMode.DisplayedCellsExceptHeaders;\\
\hspace*{0.5cm}dataGridA3.ColumnHeadersBorderStyle =
            DataGridViewHeaderBorderStyle.Raised;\\
\hspace*{0.5cm}dataGridA3.CellBorderStyle =
            DataGridViewCellBorderStyle.Single;\\
\hspace*{0.5cm}dataGridA3.GridColor =
            Color.DodgerBlue;\\
\hspace*{0.5cm}dataGridA3.ColumnHeadersVisible =
            false;\\
\hspace*{0.5cm}dataGridA3.RowHeadersVisible = false;\\
\hspace*{0.5cm}dataGridA3.BackgroundColor =
            Color.WhiteSmoke;\\
\hspace*{0.5cm}dataGridA3.BorderStyle =
            BorderStyle.None;\\
\hspace*{0.5cm}dataGridA3.AllowUserToResizeColumns =
            true;\\[0.1cm]
\hspace*{0.5cm}DataGridViewCellStyle columnHeaderStyle = new
            DataGridViewCellStyle();\\
\hspace*{0.5cm}columnHeaderStyle.BackColor =
            Color.AntiqueWhite;\\
\hspace*{0.5cm}columnHeaderStyle.Alignment =
            DataGridViewContentAlignment.MiddleCenter;\\
\hspace*{0.5cm} dataGridA3.ColumnHeadersDefaultCellStyle =
           columnHeaderStyle;\\[0.1cm]
\hspace*{0.5cm}DataGridViewCell CellR = new
            DataGridViewTextBoxCell();\\
\hspace*{0.5cm}CellR.Style.Alignment =
            DataGridViewContentAlignment.MiddleRight;\\[0.1cm]
\hspace*{0.5cm}DataGridViewCell CellL = new
        DataGridViewTextBoxCell();\\
\hspace*{0.5cm}CellL.Style.Alignment =
            DataGridViewContentAlignment.MiddleLeft;\\[0.1cm]
\hspace*{0.5cm}int line = 0;\\
\hspace*{0.5cm} line =
           Convert.ToInt16(textLineA3.Text);\\
\hspace*{0.5cm}dataGridA3.RowCount = line;\\[0.1cm]
\hspace*{0.5cm}for (int i = 0; i $<$ column; i++)\\
\hspace*{0.5cm}$\{$\\
\hspace*{1cm}dataGridA3.Columns[i].Name = "C" + (i +1);\\
\hspace*{1cm}dataGridA3.Columns[i].CellTemplate =
                CellR;\\
\hspace*{1cm}dataGridA3.Columns[i].Width = 30;\\
\hspace*{0.5cm}$\}$\\
$\}$\\
private void initMatrixScalarProduct()\\
$\{$\\
\hspace*{0.5cm}int column = 0;\\
\hspace*{0.5cm}column = Convert.ToInt16(textcolumnA3.Text);\\
\hspace*{0.5cm}dataGridScalarProduct.ColumnCount =
            column;\\
\hspace*{0.5cm}dataGridScalarProduct.AllowUserToOrderColumns =
            false;\\
\hspace*{0.5cm}dataGridScalarProduct.AllowUserToAddRows =
            false;\\
\hspace*{0.5cm}dataGridScalarProduct.Enabled = true;\\
\hspace*{0.5cm}dataGridScalarProduct.AutoSizeRowsMode =\\
            DataGridViewAutoSizeRowsMode.DisplayedCellsExceptHeaders;\\
\hspace*{0.5cm}dataGridScalarProduct.ColumnHeadersBorderStyle =
            DataGridViewHeaderBorderStyle.Raised;\\
\hspace*{0.5cm}dataGridScalarProduct.CellBorderStyle =
            DataGridViewCellBorderStyle.Single;\\
\hspace*{0.5cm}dataGridScalarProduct.GridColor =
            Color.DodgerBlue;\\
\hspace*{0.5cm}dataGridScalarProduct.ColumnHeadersVisible =
            false;\\
\hspace*{0.5cm}dataGridScalarProduct.RowHeadersVisible =
            false;\\
\hspace*{0.5cm}dataGridScalarProduct.BackgroundColor =
            Color.WhiteSmoke;\\
\hspace*{0.5cm}dataGridScalarProduct.BorderStyle =
            BorderStyle.None;\\
\hspace*{0.5cm}dataGridScalarProduct.AllowUserToResizeColumns =
            true;\\[0.1cm]
\hspace*{0.5cm}DataGridViewCellStyle columnHeaderStyle = new
            DataGridViewCellStyle();\\
\hspace*{0.5cm}columnHeaderStyle.BackColor =
            Color.AntiqueWhite;\\
\hspace*{0.5cm}columnHeaderStyle.Alignment =
            DataGridViewContentAlignment.MiddleCenter;\\
\hspace*{0.5cm}dataGridScalarProduct.ColumnHeadersDefaultCellStyle =
            columnHeaderStyle;\\[0.1cm]
\hspace*{0.5cm}DataGridViewCell CellR = new
            DataGridViewTextBoxCell();\\
\hspace*{0.5cm}CellR.Style.Alignment =
            DataGridViewContentAlignment.MiddleRight;\\[0.1cm]
\hspace*{0.5cm}DataGridViewCell CellL = new
            DataGridViewTextBoxCell();\\
\hspace*{0.5cm}CellL.Style.Alignment =
            DataGridViewContentAlignment.MiddleLeft;\\[0.1cm]
\hspace*{0.5cm}int line = 0;\\
\hspace*{0.5cm}line =
            Convert.ToInt16(textLineA3.Text);\\
\hspace*{0.5cm}dataGridScalarProduct.RowCount =
            line;\\[0.1cm]
\hspace*{0.5cm}for (int i = 0; i $<$ column; i++)\\
\hspace*{0.5cm}$\{$\\
\hspace*{1cm}dataGridScalarProduct.Columns[i].Name = "C" + (i +
                1);\\
\hspace*{1cm}dataGridScalarProduct.Columns[i].CellTemplate =
                CellR;\\
\hspace*{1cm}dataGridScalarProduct.Columns[i].Width =
                30;\\
\hspace*{0.5cm}$\}$\\
$\}$\\
private void Generating butons3$_{-}$Click(object sender, EventArgs
        e)\\
$\{$\\
\hspace*{0.5cm}if (textLineA3.Text != "")\\
\hspace*{0.5cm}$\{$\\
\hspace*{1cm}  initMatrixA3();\\
\hspace*{0.5cm}$\}$\\
\hspace*{0.5cm}else\\
\hspace*{1cm}MessageBox.Show("Introduce number of lines and number
of columns!");\\
$\}$\\
private void btScalarProduct$_{-}$Click(object sender, EventArgs e)
$\{$\\
\hspace*{0.5cm}initMatrixScalarProduct ();\\
\hspace*{0.5cm}if (textScalar.Text != "")\\
\hspace*{0.5cm}$\{$\\
\hspace*{1cm}int lineA = 0;\\
\hspace*{1cm}int columnA = 0;\\
\hspace*{1cm}columnA =
                Convert.ToInt16(textcolumnA3.Text);\\
\hspace*{1cm}lineA =
                Convert.ToInt16(textLineA3.Text);\\
\hspace*{1cm}int[,] A3 = new int[lineA, columnA];\\
\hspace*{1cm}int[,] ScalarProduct = new int[lineA,
                columnA];\\
\hspace*{1cm}int a;\\
\hspace*{1cm}$\#$region reading of scalar a and matrix
                A\\
\hspace*{1cm}a = Convert.ToInt16(textScalar.Text);\\
\hspace*{1cm}for (int i = 0; i $<$ lineA; i++)\\
\hspace*{1cm}$\{$\\
\hspace*{1.5cm}for (int j = 0; j $<$ columnA;
                    j++)\\
\hspace*{1.5cm}$\{$\\
\hspace*{1.8cm} if (dataGridA3.Rows[i].Cells[j].Value.ToString() ==
                       "E")\\
\hspace*{2cm}  A3[i, j] =
                          Int32.MinValue;\\
\hspace*{1.8cm}else\\
\hspace*{2cm}A3[i, j] =Convert.ToInt16(dataGridA3.Rows[i].Cells[j].Value.ToString());\\
\hspace*{1.5cm}$\}$\\
\hspace*{1cm}$\}$\\
\hspace*{1cm}$\#$endregion\\
\hspace*{1cm}$\#$region computation of scalar product\\
\hspace*{1cm}for (int i = 0; i $<$ lineA; i++)\\
\hspace*{1cm}$\{$\\
\hspace*{1.5cm}for (int j = 0; j $<$ columnA;
                    j++)\\
\hspace*{1.5cm}$\{$\\
\hspace*{1.8cm}if (A3[i, j] == Int32.MinValue)\\
\hspace*{2cm}Scalar Product [i, j] = Int32.MinValue;\\
\hspace*{1.8cm}else\\
\hspace*{2cm}Scalar Product [i, j] = (A3[i, j] + a);\\
\hspace*{1.5cm}$\}$\\
\hspace*{1cm}$\}$\\
\hspace*{1cm}$\#$endregion\\
\hspace*{1cm}$\#$region display scalar product\\
\hspace*{1cm}for (int i = 0; i $<$ lineA; i++)\\
 \hspace*{1cm}$\{$\\
 \hspace*{1.5cm}for (int j = 0; j $<$ columnA; j++)\\
 \hspace*{1.5cm}$\{$\\
 \hspace*{1.8cm}if (ScalarProduct [i, j] == Int32.MinValue)\\
 \hspace*{2cm}dataGridScalarProduct.Rows[i].Cells[j].Value = "E";\\
 \hspace*{1.8cm}else\\
 \hspace*{2cm}dataGridScalarProduct.Rows[i].Cells[j].Value = ScalarProduct[i,j];\\
 \hspace*{1.5cm}$\}$\\
 \hspace*{1cm}$\}$\\
 \hspace*{1cm}$\#$endregion\\
 \hspace*{0.5cm}$\}$\\
$\}$\\
public void Reset$_{-}$Values$_{-}$for$_{-}$Scalar Multiplication()\\
$\{$\\
\hspace*{0.5cm}textcolumnA3.ResetText();\\
\hspace*{0.5cm}textScalar.ResetText();\\
\hspace*{0.5cm}textLineA3.ResetText();\\
\hspace*{0.5cm}dataGridA3.Rows.Clear();\\
\hspace*{0.5cm}dataGridScalarProduct.Rows.Clear();\\
$\}$\\[0.1cm]
private void btReset$_{-}$Scalar$_{-}$Product$_{-}$Click(object
sender, EventArgs e)\\
$\{$\\
\hspace*{0.5cm}Reset$_{-}$Values$_{-}$for$_{-}$Scalar Multiplication();\\
$\}$\\
$\#$endregion\\[0.2cm]
$\#$region {\it Power of a  matrix in max plus algebra}\\[0.1cm]
private void initMatrixA4()
$\{$\\
\hspace*{0.5cm}int column = 0;\\
\hspace*{0.5cm}column =
            Convert.ToInt16(textlineA4.Text);\\
\hspace*{0.5cm}dataGridA4.ColumnCount = column;\\
\hspace*{0.5cm}dataGridA4.AllowUserToOrderColumns =
            false;\\
\hspace*{0.5cm}dataGridA4.AllowUserToAddRows = false;\\
\hspace*{0.5cm}dataGridA4.Enabled = true;\\
\hspace*{0.5cm}dataGridA4.AutoSizeRowsMode =\\
            DataGridViewAutoSizeRowsMode.DisplayedCellsExceptHeaders;\\
\hspace*{0.5cm}dataGridA4.ColumnHeadersBorderStyle =
            DataGridViewHeaderBorderStyle.Raised;\\
\hspace*{0.5cm}dataGridA4.CellBorderStyle =
            DataGridViewCellBorderStyle.Single;\\
\hspace*{0.5cm}dataGridA4.GridColor =
            Color.DodgerBlue;\\
\hspace*{0.5cm}dataGridA4.ColumnHeadersVisible =
            false;\\
\hspace*{0.5cm}dataGridA4.RowHeadersVisible = false;\\
\hspace*{0.5cm}dataGridA4.BackgroundColor =
            Color.WhiteSmoke;\\
\hspace*{0.5cm}dataGridA4.BorderStyle =
            BorderStyle.None;\\
\hspace*{0.5cm}dataGridA4.AllowUserToResizeColumns =
            true;\\[0.1cm]
\hspace*{0.5cm}DataGridViewCellStyle columnHeaderStyle = new
DataGridViewCellStyle();\\
\hspace*{0.5cm}columnHeaderStyle.BackColor =
            Color.AntiqueWhite;\\
\hspace*{0.5cm}columnHeaderStyle.Alignment =
            DataGridViewContentAlignment.MiddleCenter;\\
\hspace*{0.5cm}dataGridA4.ColumnHeadersDefaultCellStyle =
            columnHeaderStyle;\\[0.1cm]
\hspace*{0.5cm}DataGridViewCell CellR = new
DataGridViewTextBoxCell();\\
\hspace*{0.5cm}CellR.Style.Alignment =
            DataGridViewContentAlignment.MiddleRight;\\[0.1cm]
\hspace*{0.5cm}DataGridViewCell CellL = new
            DataGridViewTextBoxCell();\\
\hspace*{0.5cm}CellL.Style.Alignment =
            DataGridViewContentAlignment.MiddleLeft;\\[0.1cm]
\hspace*{0.5cm}int line = 0;\\
\hspace*{0.5cm}line =
            Convert.ToInt16(textlineA4.Text);\\
\hspace*{0.5cm}dataGridA4.RowCount = line;\\[0.1cm]
\hspace*{0.5cm}for (int i = 0; i $<$ column; i++)\\
\hspace*{0.5cm}$\{$\\
\hspace*{1cm}dataGridA4.Columns[i].Name = "C" + (i + 1);\\
\hspace*{1cm}dataGridA4.Columns[i].CellTemplate = CellR;\\
\hspace*{0.5cm}dataGridA4.Columns[i].Width = 30;\\
\hspace*{0.5cm}$\}$\\
$\}$\\
private void initPowerMatrix ()\\
\hspace*{0.5cm}$\{$\\
\hspace*{0.5cm}int column = 0;\\
\hspace*{0.5cm}column = Convert.ToInt16(textlineA4.Text);\\
\hspace*{0.5cm}dataGridMatrix$_{-}$at$_{-}$power$_{-}$n.ColumnCount = column;\\
\hspace*{0.5cm}dataGridMatrix$_{-}$at$_{-}$power$_{-}$n.AllowUserToOrderColumns=false;\\
\hspace*{0.5cm}dataGridMatrix$_{-}$at$_{-}$power$_{-}$n.AllowUserToAddRows=false;\\
\hspace*{0.5cm}dataGridMatrix$_{-}$at$_{-}$power$_{-}$n.Enabled =true;\\
\hspace*{0.5cm}dataGridMatrix$_{-}$at$_{-}$power$_{-}$n.AutoSizeRowsMode=\\
DataGridViewAutoSizeRowsMode.DisplayedCellsExceptHeaders;\\
\hspace*{0.5cm}dataGridMatrix$_{-}$at$_{-}$power$_{-}$n.ColumnHeadersBorderStyle
=DataGridViewHeaderBorderStyle.Raised;\\
\hspace*{0.5cm}dataGridMatrix$_{-}$at$_{-}$power$_{-}$n.CellBorderStyle
=DataGridViewCellBorderStyle.Single;\\
\hspace*{0.5cm}dataGridMatrix$_{-}$at$_{-}$power$_{-}$n.GridColor =Color.DodgerBlue;\\
\hspace*{0.5cm}dataGridMatrix$_{-}$at$_{-}$power$_{-}$n.ColumnHeadersVisible=false;\\
\hspace*{0.5cm}dataGridMatrix$_{-}$at$_{-}$power$_{-}$n.RowHeadersVisible=false;\\
\hspace*{0.5cm}dataGridMatrix$_{-}$at$_{-}$power$_{-}$n.BackgroundColor
=Color.WhiteSmoke;\\
\hspace*{0.5cm}dataGridMatrix$_{-}$at$_{-}$power$_{-}$n.BorderStyle
=BorderStyle.None;\\
\hspace*{0.5cm}dataGridMatrix$_{-}$at$_{-}$power$_{-}$n.AllowUserToResizeColumns=true;\\[0.1cm]
\hspace*{0.5cm}DataGridViewCellStyle columnHeaderStyle = new
DataGridViewCellStyle();\\
\hspace*{0.5cm}columnHeaderStyle.BackColor = Color.AntiqueWhite;\\
\hspace*{0.5cm}columnHeaderStyle.Alignment =
            DataGridViewContentAlignment.MiddleCenter;\\
\hspace*{0.5cm}dataGridMatrix$_{-}$at$_{-}$power$_{-}$n.ColumnHeadersDefaultCellStyle
=columnHeaderStyle;\\[0.1cm]
\hspace*{0.5cm}DataGridViewCell CellR = new
            DataGridViewTextBoxCell();\\
\hspace*{0.5cm}CellR.Style.Alignment =
            DataGridViewContentAlignment.MiddleRight;\\[0.1cm]
\hspace*{0.5cm}DataGridViewCell CellL = new
            DataGridViewTextBoxCell();\\
\hspace*{0.5cm}CellL.Style.Alignment =
            DataGridViewContentAlignment.MiddleLeft;\\[0.1cm]
\hspace*{0.5cm}int line = 0;\\
\hspace*{0.5cm}line =
            Convert.ToInt16(textlinieA4.Text);\\
\hspace*{0.5cm}dataGridMatrix$_{-}$at$_{-}$power$_{-}$n.RowCount =
            line;\\[0.1cm]
\hspace*{0.5cm}for (int i = 0; i $<$ column; i++)\\
\hspace*{0.5cm}$\{$\\
\hspace*{1cm}dataGridMatrix$_{-}$at$_{-}$power$_{-}$n.Columns[i].Name
= "C" + (i +1);\\
\hspace*{1cm}dataGridMatrix$_{-}$at$_{-}$power$_{-}$n.Columns[i].CellTemplate
=CellR;\\
\hspace*{1cm}dataGridMatrix$_{-}$at$_{-}$power$_{-}$n.Columns[i].Width
=30;\\
\hspace*{0.5cm}$\}$\\
$\}$\\
private void btGenerareA4$_{-}$Click(object sender, EventArgs e)\\
$\{$\\
\hspace*{0.5cm}if (textlineA4.Text != "")\\
\hspace*{0.5cm}$\{$\\
\hspace*{1cm}initMatrixA4();\\
\hspace*{0.5cm}$\}$\\
\hspace*{0.5cm}else\\
\hspace*{1cm}MessageBox.Show("Introduce number of lines and
columns!");
$\}$\\
private void btComputPower$_{-}$Click(object sender, EventArgs e)\\
$\{$\\
\hspace*{0.5cm}initMatrixPower ();\\
\hspace*{0.5cm}labelPower.Text = textPower.Text;\\
\hspace*{0.5cm}if (textPower.Text != "")\\
\hspace*{0.5cm}$\{$\\
\hspace*{1cm}int lineA = 0;\\
\hspace*{1cm}int columnA = 0;\\
\hspace*{1cm}columnA =
                Convert.ToInt16(textlineA4.Text);\\
\hspace*{1cm}lineA =
                Convert.ToInt16(textlineA4.Text);\\
\hspace*{1cm}int[,] A4 = new int[lineA, columnA];\\
\hspace*{1cm}int[,] B = new int[lineA, columnA];\\
\hspace*{1cm}int[,] Power$_{-}$n = new int[lineA,
                columnA];\\
\hspace*{1cm}int[,] sum = new int[lineA, columnA];\\
\hspace*{1cm}int a;\\
\hspace*{1cm}a = Convert.ToInt16(textPower.Text);\\
\hspace*{1cm}$\#$region reading matrix A\\
\hspace*{1cm}for (int i = 0; i $<$ lineA; i++)\\
\hspace*{1cm}$\{$\\
\hspace*{1.5cm}for (int j = 0; j $<$ columnA;j++)\\
\hspace*{1.5cm}$\{$\\
\hspace*{1.5cm}if (dataGridA4.Rows[i].Cells[j].Value.ToString() == "E")\\
\hspace*{1.8cm}  A4[i, j] =Int32.MinValue;\\
\hspace*{1.5cm}else\\
\hspace*{1.8cm}A4[i, j] =
Convert.ToInt16(dataGridA4.Rows[i].Cells[j].Value.ToString());\\
\hspace*{1.5cm}B[i, j] = A4[i, j];\\
\hspace*{1.5cm}$\}$\\
\hspace*{1cm}$\}$\\
\hspace*{1cm}$\#$endregion\\
\hspace*{1cm}$\#$region PowerMatrix A\\
\hspace*{1cm}for (int p = 2; p $<=$ a; p++)\\
\hspace*{1cm}$\{$\\
\hspace*{1.5cm}for (int i = 0; i $<$ lineA;i++)\\
\hspace*{1.5cm}$\{$\\
\hspace*{1.5cm} for (int j = 0; j $<$ columnA;j++)\\
\hspace*{1.5cm}$\{$\\
\hspace*{1.8cm}Power$_{-}$n[i, j] = Int32.MinValue;\\
\hspace*{1.8cm}for (int k = 0; k $<$ lineA; k++)\\
\hspace*{1.8cm}$\{$\\
 \hspace*{1.8cm}if (A4[i, k] == Int32.MinValue || B[k, j] ==
 Int32.MinValue)\\
 \hspace*{2cm}sum[i, j] =Int32.MinValue;\\
 \hspace*{1.8cm}else\\
 \hspace*{2cm}sum[i, j] = A4[i, k] + B[k,j];\\[0.1cm]
  \hspace*{1.8cm}if (Power$_{-}$n[i, j] $>$ sum[i,j])\\
  \hspace*{2cm}Power$_{-}$n[i, j] = Power$_{-}$n[i,j];\\
  \hspace*{1.8cm}else\\
  \hspace*{2cm}Power$_{-}n$[i, j] = sum[i,j];\\[0.1cm]
  \hspace*{1.8cm}$\}$\\
  \hspace*{1.5cm}$\}$\\
   \hspace*{1cm}$\}$\\
   \hspace*{1cm}for (int i = 0; i $<$ lineA;i++)\\
   \hspace*{1cm}$\{$\\
   \hspace*{1.5cm}for (int j = 0; j $<$ columnA;j++)\\
   \hspace*{1.5cm}$\{$\\
   \hspace*{1.5cm}B[i, j] = Power$_{-}$n[i,j];\\
 \hspace*{1cm}$\}$\\
 \hspace*{0.5cm}$\}$\\
 $\}$\\
 $\#$endregion\\
 $\#$region dysplay matrix\\
 for (int i = 0; i $<$ lineA; i++)\\
 $\{$\\
 \hspace*{0.5cm}for (int j = 0; j $<$ columnA;j++)\\
 \hspace*{0.5cm}$\{$\\
 \hspace*{1cm}if (Power$_{-}$n[i, j] ==
                        Int32.MinValue)\\
 \hspace*{1cm}dataGridMatrix$_{-}$at$_{-}$power$_{-}$n.Rows[i].Cells[j].Value ="E";\\
 \hspace*{0.5cm}else\\
 \hspace*{1cm}dataGridMatrix$_{-}$at$_{-}$Power$_{-}$n.Rows[i].Cells[j].Value = Power$_{-}$n[i,j];\\
 \hspace*{0.5cm}$\}$\\
 \hspace*{0.5cm}$\}$\\
 \hspace*{0.5cm}$\#$endregion\\
 \hspace*{0.5cm}$\}$\\
 $\}$\\
 public void Reset$_{-}$Values$_{-}$for$_{-}$Lifting$_{-}$at$_{-}$Power()\\
 $\{$\\
 \hspace*{0.5cm}textlineA4.ResetText();\\
 \hspace*{0.5cm}textPower.ResetText();\\
 \hspace*{0.5cm}dataGridA4.Rows.Clear();\\
 \hspace*{0.5cm}dataGridMatrix$_{-}$at$_{-}$Power$_{-}$n.Rows.Clear();\\
 $\}$\\
 private void btReset$_{-}$lifting$_{-}$at$_{-}$power$_{-}$Click(object sender, EventArgs e)\\
 $\{$\\
 \hspace*{0.5cm}Reset$_{-}$Values$_{-}$for$_{-}$Lifting$_{-}$at$_{-}$Power();\\
 $\}$\\
$\#$endregion\\[0.2cm]
$\#$region {\it Solving equations system in max plus algebra}\\[0.1cm]
private void initMatrixA5()\\
$\{$\\
\hspace*{0.5cm}int column = 0;\\
\hspace*{0.5cm}column = Convert.ToInt16(textlinieA6.Text);\\
\hspace*{0.5cm}dataGridA5.ColumnCount = column;\\
\hspace*{0.5cm}dataGridA5.AllowUserToOrderColumns =false;\\
\hspace*{0.5cm}dataGridA5.AllowUserToAddRows = false;\\
\hspace*{0.5cm}dataGridA5.Enabled = true;\\
\hspace*{0.5cm}dataGridA5.AutoSizeRowsMode =\\
 DataGridViewAutoSizeRowsMode.DisplayedCellsExceptHeaders;\\
\hspace*{0.5cm}dataGridA5.ColumnHeadersBorderStyle =
 DataGridViewHeaderBorderStyle.Raised;\\
\hspace*{0.5cm}dataGridA5.CellBorderStyle =
 DataGridViewCellBorderStyle.Single;\\
\hspace*{0.5cm}dataGridA5.GridColor = Color.DodgerBlue;\\
\hspace*{0.5cm}dataGridA5.ColumnHeadersVisible =false;\\
\hspace*{0.5cm}dataGridA5.RowHeadersVisible = false;\\
\hspace*{0.5cm}dataGridA5.BackgroundColor = Color.WhiteSmoke;\\
\hspace*{0.5cm}dataGridA5.BorderStyle = BorderStyle.None;\\
\hspace*{0.5cm}dataGridA5.AllowUserToResizeColumns =true;\\[0.1cm]
\hspace*{0.5cm}DataGridViewCellStyle columnHeaderStyle = new
 DataGridViewCellStyle();\\
\hspace*{0.5cm}columnHeaderStyle.BackColor = Color.AntiqueWhite;\\
\hspace*{0.5cm}columnHeaderStyle.Alignment =
 DataGridViewContentAlignment.MiddleCenter;\\
\hspace*{0.5cm}dataGridA5.ColumnHeadersDefaultCellStyle =
 columnHeaderStyle;\\[0.1cm]
\hspace*{0.5cm}DataGridViewCell CellR = new
 DataGridViewTextBoxCell();\\
\hspace*{0.5cm}CellR.Style.Alignment =
 DataGridViewContentAlignment.MiddleRight;\\[0.1cm]
\hspace*{0.5cm}DataGridViewCell CellL = new
 DataGridViewTextBoxCell();\\
\hspace*{0.5cm}CellL.Style.Alignment =
 DataGridViewContentAlignment.MiddleLeft;\\[0.1cm]
\hspace*{0.5cm}int line = 0;\\
\hspace*{0.5cm}line = Convert.ToInt16(textlineA5.Text);\\
\hspace*{0.5cm}dataGridA5.RowCount = line;\\[0.1cm]
\hspace*{0.5cm}for (int i = 0; i $<$ column; i++)\\
\hspace*{0.5cm}$\{$\\
\hspace*{1cm}dataGridA5.Columns[i].Name = "C" + (i +1);\\
\hspace*{1cm}dataGridA5.Columns[i].CellTemplate =CellR;\\
\hspace*{1cm}dataGridA5.Columns[i].Width = 30;\\
\hspace*{0.5cm}$\}$\\
$\}$\\
private void initMatrixX0()\\
$\{$\\
\hspace*{0.5cm}int column = 0;\\
\hspace*{0.5cm}column = 1;\\
\hspace*{0.5cm}dataGridX0.ColumnCount = column;\\
\hspace*{0.5cm}dataGridX0.AllowUserToOrderColumns = false;\\
\hspace*{0.5cm}dataGridX0.AllowUserToAddRows = false;\\
\hspace*{0.5cm}dataGridX0.Enabled = true;\\
\hspace*{0.5cm}dataGridX0.AutoSizeRowsMode =\\
\hspace*{0.5cm}DataGridViewAutoSizeRowsMode.DisplayedCellsExceptHeaders;\\
\hspace*{0.5cm}dataGridX0.ColumnHeadersBorderStyle =
DataGridViewHeaderBorderStyle.Raised;\\
\hspace*{0.5cm}dataGridX0.CellBorderStyle =
 DataGridViewCellBorderStyle.Single;\\
\hspace*{0.5cm}dataGridX0.GridColor = Color.DodgerBlue;\\
\hspace*{0.5cm}dataGridX0.ColumnHeadersVisible = false;\\
\hspace*{0.5cm}dataGridX0.RowHeadersVisible = false;\\
\hspace*{0.5cm}dataGridX0.BackgroundColor = Color.WhiteSmoke;\\
\hspace*{0.5cm}dataGridX0.BorderStyle = BorderStyle.None;\\
\hspace*{0.5cm}dataGridX0.AllowUserToResizeColumns = true;\\[0.1cm]
\hspace*{0.5cm}DataGridViewCellStyle columnHeaderStyle = new
 DataGridViewCellStyle();\\
\hspace*{0.5cm}columnHeaderStyle.BackColor = Color.AntiqueWhite;\\
\hspace*{0.5cm}columnHeaderStyle.Alignment =
 DataGridViewContentAlignment.MiddleCenter;\\
 \hspace*{0.5cm}dataGridX0.ColumnHeadersDefaultCellStyle =
  columnHeaderStyle;\\[0.1cm]
\hspace*{0.5cm}DataGridViewCell CellR = new
  DataGridViewTextBoxCell();\\
\hspace*{0.5cm}CellR.Style.Alignment =
  DataGridViewContentAlignment.MiddleRight;\\[0.1cm]
\hspace*{0.5cm}DataGridViewCell CellL = new
  DataGridViewTextBoxCell();\\
\hspace*{0.5cm}CellL.Style.Alignment =
  DataGridViewContentAlignment.MiddleLeft;\\[0.1cm]
\hspace*{0.5cm}int line = 0;\\
\hspace*{0.5cm}line = Convert.ToInt16(textlineA5.Text);\\
\hspace*{0.5cm}dataGridX0.RowCount = line;\\[0.1cm]
\hspace*{0.5cm}for (int i = 0; i $<$ column; i++)\\
\hspace*{0.5cm}$\{$\\
\hspace*{1cm}dataGridX0.Columns[i].Name = "C" + (i + 1);\\
\hspace*{1cm}dataGridX0.Columns[i].CellTemplate = CellR;\\
\hspace*{1cm}dataGridX0.Columns[i].Width = 30;\\
\hspace*{0.5cm}$\}$\\
$\}$\\
private void initMatrixSolutionSystem()\\
$\{$\\
\hspace*{0.5cm}int column = 0;\\
\hspace*{0.5cm}column = 1;\\
\hspace*{0.5cm}dataGridSolutionXk.ColumnCount = column;\\
\hspace*{0.5cm}dataGridSolutionXk.AllowUserToOrderColumns = false;\\
\hspace*{0.5cm}dataGridSolutionXk.AllowUserToAddRows = false;\\
\hspace*{0.5cm}dataGridSolutionXk.Enabled = true;\\
\hspace*{0.5cm}dataGridSolutionXk.AutoSizeRowsMode =\\
\hspace*{0.5cm}DataGridViewAutoSizeRowsMode.DisplayedCellsExceptHeaders;\\
\hspace*{0.5cm}dataGridSolutionXk.ColumnHeadersBorderStyle =
\hspace*{0.5cm}DataGridViewHeaderBorderStyle.Raised;\\
\hspace*{0.5cm}dataGridSolutionXk.CellBorderStyle =
\hspace*{0.5cm}DataGridViewCellBorderStyle.Single;\\
\hspace*{0.5cm}dataGridSolutionXk.GridColor = Color.DodgerBlue;\\
\hspace*{0.5cm}dataGridSolutionXk.ColumnHeadersVisible = false;\\
\hspace*{0.5cm}dataGridSolutionXk.RowHeadersVisible = false;\\
\hspace*{0.5cm}dataGridSolutionXk.BackgroundColor = Color.WhiteSmoke;\\
\hspace*{0.5cm}dataGridSolutionXk.BorderStyle = BorderStyle.None;\\
\hspace*{0.5cm}dataGridSolutionXk.AllowUserToResizeColumns = true;\\[0.1cm]
\hspace*{0.5cm}DataGridViewCellStyle columnHeaderStyle = new
DataGridViewCellStyle();\\
\hspace*{0.5cm}columnHeaderStyle.BackColor = Color.AntiqueWhite;\\
\hspace*{0.5cm}columnHeaderStyle.Alignment =
  DataGridViewContentAlignment.MiddleCenter;\\
\hspace*{0.5cm}dataGridSolutieXk.ColumnHeadersDefaultCellStyle =
  columnHeaderStyle;\\[0.1cm]
\hspace*{0.5cm}DataGridViewCell CellR = new
  DataGridViewTextBoxCell();\\
 \hspace*{0.5cm}CellR.Style.Alignment =
  DataGridViewContentAlignment.MiddleRight;\\[0.1cm]
\hspace*{0.5cm}DataGridViewCell CellL = new
 DataGridViewTextBoxCell();\\
 \hspace*{0.5cm}CellL.Style.Alignment =
  DataGridViewContentAlignment.MiddleLeft;\\[0.1cm]
\hspace*{0.5cm}int line = 0;\\
\hspace*{0.5cm}line = Convert.ToInt16(textlineA5.Text);\\
 \hspace*{0.5cm}dataGridSolutionXk.RowCount = line;\\[0.1cm]
 \hspace*{0.5cm}for (int i = 0; i $<$ column; i++)\\
 \hspace*{0.5cm}$\{$\\
 \hspace*{1cm} dataGridSolutionXk.Columns[i].Name = "C" + (i + 1);\\
 \hspace*{1cm}dataGridSolutionXk.Columns[i].CellTemplate = CellR;\\
 \hspace*{1cm}dataGridSolutionXk.Columns[i].Width = 30;\\[0.1cm]
 \hspace*{0.5cm}$\}$\\
 $\}$\\
 private void btGenerateMatrices$_{-}$Click(object sender, EventArgs e)\\
 $\{$\\
 \hspace*{0.5cm}if (textlineA5.Text != "")\\
 \hspace*{0.5cm}$\{$\\
 \hspace*{1cm}initMatrixA5();\\
 \hspace*{1cm}initMatrixX0();\\
 \hspace*{0.5cm}$\}$\\
 \hspace*{0.5cm}else\\
 \hspace*{1cm}MessageBox.Show("Introduce number of lines and columns!");\\[0.1cm]
 $\}$\\
 private void btComputSystem$_{-}$Click(object sender, EventArgs e)\\
 $\{$\\
 \hspace*{0.5cm} initMatrixSolutionSystem();\\
 \hspace*{0.5cm}label$_{-}$k.Text = textk.Text;\\
 \hspace*{0.5cm}if (textk.Text != "")\\
 \hspace*{0.5cm}$\{$\\
 \hspace*{1cm}int lineA = 0;\\
 \hspace*{1cm}int columnA = 0;\\
 \hspace*{1cm}int linex0 = 0;\\
 \hspace*{1cm} columnA =
  Convert.ToInt16(textlinieA5.Text);\\
 \hspace*{1cm}lineA =
  Convert.ToInt16(textlinieA5.Text);\\
 \hspace*{1cm}linex0 = Convert.ToInt16(textlineA5.Text);\\
 \hspace*{1cm}\hspace*{1cm}int[,] A5 = new int[lineA,columnA];\\
 \hspace*{1cm}nt[,] X0 = new int[linex0, 1];\\
 \hspace*{1cm}int[,] Xk = new int[lineA, 1];\\
 \hspace*{1cm}int[,] B = new int[lineA, columnA];\\
 \hspace*{1cm}int[,] Power$_{-}$n = new int[lineA, columnA];\\
 \hspace*{1cm}int[,] sum = new int[lineA, columnA];\\
 \hspace*{1cm}int[,] sum2 = new int[lineA, 1];\\
 \hspace*{1cm}int k;\\
 \hspace*{1cm}k = Convert.ToInt16(textk.Text);\\
 \hspace*{0.5cm}$\#$region readings matrix A and matrix X0\\
 \hspace*{0.5cm}for (int i = 0; i $<$ lineA; i++)\\
 \hspace*{0.5cm}$\{$\\
 \hspace*{1cm}for (int j = 0; j $<$ coloanaA; j++)\\
 \hspace*{1cm}$\{$\\
 \hspace*{1.5cm}if (dataGridA5.Rows[i].Cells[j].Value.ToString() == "E")\\
 \hspace*{1.8cm}A5[i, j] =Int32.MinValue;\\
 \hspace*{1.5cm}else\\
 \hspace*{1.8cm}A5[i, j] =
  Convert.ToInt16(dataGridA5.Rows[i].Cells[j].Value.ToString());\\
  \hspace*{1.5cm}B[i, j] = A5[i, j];\\
  \hspace*{1cm}$\}$\\
  \hspace*{0.5cm}$\}$
  \hspace*{0.5cm}for (int i = 0; i $<$ linex0; i++)\\
  \hspace*{0.5cm}$\{$\\
\hspace*{1cm}if (dataGridX0.Rows[i].Cells[0].Value.ToString() == "E")\\
\hspace*{1.5cm}X0[i, 0] = Int32.MinValue;\\
\hspace*{1cm}else\\
\hspace*{1.5cm}X0[i, 0] =
 Convert.ToInt16(dataGridX0.Rows[i].Cells[0].Value.ToString());\\
 \hspace*{0.5cm}$\}$\\
 \hspace*{0.5cm}$\#$endregion\\
 \hspace*{0.5cm}$\#$region lifting at power k of matrix A\\
 \hspace*{0.5cm}if (k == 1)\\
 \hspace*{0.5cm}$\{$\\
 \hspace*{1cm}for (int i = 0; i $<$ lineA; i++)\\
 \hspace*{1cm}$\{$\\
 \hspace*{1.5cm}for (int j = 0; j $<$ columnA; j++)\\
 \hspace*{1.5cm}$\{$\\
 \hspace*{1.8cm}Power$_{-}$n[i, j] = A5[i,j];\\
 \hspace*{1.5cm}$\}$\\
 \hspace*{1cm}$\}$\\
 \hspace*{0.5cm}$\}$\\
 \hspace*{0.5cm}else\\
 \hspace*{0.5cm}$\{$\\
  \hspace*{1cm}for (int p = 2; p $<=$ k; p++)\\
  \hspace*{1cm}$\{$\\
 \hspace*{1.5cm}for (int i = 0; i $<$ lineA; i++)\\
 \hspace*{1.5cm}$\{$\\
 \hspace*{1.5cm}for (int j = 0; j $<$ columnA; j++)\\
 \hspace*{1.5cm}$\{$\\
\hspace*{1.8cm} Power$_{-}$n[i, j] = Int32.MinValue;\\
\hspace*{1.8cm}for (int h = 0; h $<$ lineA; h++)\\
\hspace*{1.8cm}$\{$\\
\hspace*{1.8cm}if (A5[i, h] == Int32.MinValue $||$ B[h, j] == Int32.MinValue)\\
\hspace*{2cm}sum[i, j] = Int32.MinValue;\\
\hspace*{1.8cm}else\\
\hspace*{2cm}sum[i, j] = A5[i, h] + B[h,j];\\
\hspace*{1.8cm} if (Power$_{-}$n[i, j] $>$ sum[i,j])\\
\hspace*{2cm}  Power$_{-}$n[i, j] = Power$_{-}$n[i,j];\\
\hspace*{1.8cm}else\\
\hspace*{2cm}Power$_{-}$n[i, j] = sum[i,j];\\
\hspace*{1.5cm}$\}$\\
\hspace*{1cm}$\}$\\
\hspace*{0.5cm}$\}$\\
\hspace*{1cm}for (int i = 0; i $<$ lineA; i++)\\
\hspace*{1.5cm}$\{$\\
\hspace*{1.8cm}for (int j = 0; j $<$ columnA;j++)\\
\hspace*{1.5cm}$\{$\\
\hspace*{1.8cm}B[i, j] = Power$_{-}$n[i, j];\\
\hspace*{1.5cm}$\}$\\
\hspace*{1cm}$\}$\\
\hspace*{0.5cm}$\}$\\
$\}$\\
$\#$endregion\\
$\#$region computation of matrix X(k)\\
for (int i = 0; i $<$ lineA; i++)\\
$\{$\\
\hspace*{0.5cm}for (int j = 0; j $<$ 1; j++)\\
\hspace*{0.5cm}$\{$\\
\hspace*{0.5cm} Xk[i, j] = Int32.MinValue;\\
\hspace*{0.5cm}for (int h = 0; h $<$ lineA; h++)\\
\hspace*{0.5cm}$\{$\\
\hspace*{1cm}if (Power$_{-}$n[i, h] == Int32.MinValue $||$ X0[h,
j]
 == Int32.MinValue)\\
 \hspace*{1.5cm}sum2[i, j] = Int32.MinValue;\\
 \hspace*{1cm}else\\
 \hspace*{1.5cm}sum2[i, j] = Power$_{-}$n[i, h] + X0[h, j];\\[0.1cm]
 \hspace*{1.5cm} if (Xk[i, j] $>$ sum2[i,j])\\
 \hspace*{1.5cm}Xk[i, j] = Xk[i,j];\\
 \hspace*{1cm}else\\
 \hspace*{1.5cm}Xk[i, j] = sum2[i,j];\\
 \hspace*{1cm}$\}$\\
 \hspace*{0.5cm}$\}$\\
 $\}$\\
 $\#$endregion\\
 \hspace*{0.5cm}$\#$region display of matrix X(k)\\
 \hspace*{0.5cm} for (int i = 0; i $<$ lineA; i++)\\
 \hspace*{0.5cm}$\{$\\
 \hspace*{1cm}if (Xk[i, 0] == Int32.MinValue)\\
 \hspace*{1.5cm}dataGridSolutionXk.Rows[i].Cells[0].Value = "E";\\
 \hspace*{1cm}else\\
 \hspace*{1.5cm}dataGridSolutionXk.Rows[i].Cells[0].Value = Xk[i,0];\\
 \hspace*{0.5cm}$\}$\\
 $\#$endregion\\
$\}$\\
$\}$\\
public void Reset$_{-}$Values$_{-}$for$_{-}$Computation$_{-}$system()\\
$\{$\\
\hspace*{0.5cm}textlineA5.ResetText();\\
\hspace*{0.5cm}textk.ResetText();\\
\hspace*{0.5cm}label$_{-}$k.Text = "k";\\
\hspace*{0.5cm}dataGridSolutionXk.Rows.Clear();\\
\hspace*{0.5cm}dataGridA5.Rows.Clear();\\
\hspace*{0.5cm}dataGridX0.Rows.Clear();\\[0.1cm]
$\}$\\
private void btResetSystem$_{-}$Click(object sender, EventArgs e)\\
$\{$\\
\hspace*{0.5cm}Reset$_{-}$Values$_{-}$for$_{-}$Computation$_{-}$System();\\
\hspace*{0.5cm}$\}$\\
\hspace*{0.5cm}$\#$endregion\\
\hspace*{0.5cm}$\}$\\
$\}$\\[0.1cm]

 We illustrate the utilization of the above programs in the
following
cases.\\[0.2cm]
{\bf 1. Sum of two matrices $A$ and $B$ in ${\bf R}_{max}$\\[0.4cm]
{\it Inputs data}\hspace*{7.5cm}{\it Outputs data}\\[0.2cm]
Number of lines $~~~~~~\fbox{3}$\\[0.1cm]
Number of columns $~\fbox{4}$\\[0.1cm]
Matrix $~A$\hspace*{2.2cm} Matrix $~B$\hspace*{3.2cm} Matrix
$~~A\oplus B$  \\[0.2cm]
 $\begin{array}{|c|c|c|c|}\hline
 3     & E & 8 & -2\cr \hline
  6    & 0 & 4 & -9\cr \hline
E    & 5 & -7 & 1\cr \hline
\end{array} ~~~~~~~~ \begin{array}{|c|c|c|c|}\hline
 9     & 9 & -1 & -5\cr \hline
  2    & -1 & 6 & -3\cr \hline
1    & 2 & 4 & -5\cr \hline
\end{array}~~~~~~~~~~~~~~~~\begin{array}{|c|c|c|c|}\hline
 9     & 9 & 8 & -2\cr \hline
  6    & 0 & 6 & -3\cr \hline
1    & 5 & 4 & 1\cr \hline
\end{array}.$\\[0.5cm]
{\bf 2. Product of two matrices $A$ and $B$ in ${\bf R}_{max}$\\[0.4cm]
{\it Inputs data}\\[0.2cm]
Number of lines of $~A!~~~~~~~ \fbox{3}~~~~~$ Number of lines of $~B!~~~~~~~~~~~~~~~~~~~ \fbox{4}$\\[0.2cm]
Number of columns of $~A!~~ \fbox{4}~~~~~~$ Number of columns of matrix $~B!~~~~ \fbox{3}$\\[0.2cm]
Matrix $A$\hspace*{2.5cm} Matrix $B$\\[0.2cm]
 $\begin{array}{|c|c|c|c|}\hline
 2     & 1 & -1 & 4\cr \hline
  E    & 0 & 5 & -3\cr \hline
-4   & -2 & E & 6\cr \hline
\end{array} ~~~~~~~~ \begin{array}{|c|c|c|}\hline
 5     & 0 & 1 \cr \hline
  7    & 4 & E \cr \hline
-5   & 9 & 2 \cr \hline
8   & -6 & 1 \cr \hline
\end{array}$\\[0.4cm]
{\it Outputs data}\\[0.2cm]
Matrix $~~~A\otimes B$\\[0.2cm]
$\begin{array}{|c|c|c|}\hline
 12     & 8 & 5 \cr \hline
  7    & 14 & 7 \cr \hline
14    & 2 & 7 \cr \hline
\end{array}.$\\[0.5cm]
{\bf 3. Multiplication with scalar $a$ of a matrix $A$ in ${\bf R}_{max}$\\[0.4cm]
{\it Inputs data}\hspace*{7.5cm}{\it Outputs data}\\[0.2cm]
Number of lines for matrix  $A!~~~~~~~~\fbox{3}$\\[0.1cm]
Number of columns for matrix $A!~~~\fbox{5}$\\[0.1cm]
Scalar $a~~~\fbox{-4}$\\[0.1cm]
Matrix $~A$\hspace*{7.5cm} Matrix $~~a\otimes
A$\\[0.2cm]
 $\begin{array}{|c|c|c|c|c|}\hline
 4     & -7 & 8 & 2 & E\cr \hline
  5    & E & 0 & E & 8\cr \hline
9    & 2 & E & 3 & 1\cr \hline
\end{array},~~~~~~~~~~~~~~~~~~~~~~~~~~~~~~~~~~~~\begin{array}{|c|c|c|c|c|}\hline
 0     & -11 & 4 & -2 & E\cr \hline
  1    & E & -4 & E & 4\cr \hline
5    & -2 & E & -1 & -3\cr \hline
\end{array}. $\\[0.5cm]
{\bf 4. Power of a matrix $A$ in ${\bf R}_{max}$\\[0.4cm]
{\it Inputs data}\hspace*{9cm}{\it Outputs data}\\[0.2cm]
Number of lines and columns for matrix $A!~~~\fbox{5}$\\[0.1cm]
Power of matrix $A!~~~\fbox{9}$\\[0.2cm]
Matrix $~A$\hspace*{9cm} Matrix $A^{(9)}$\\[0.2cm]
 $\begin{array}{|c|c|c|c|c|}\hline
 1  & 0 & -2 & E & 3\cr \hline
 0  & 2 & E & 4 & 1\cr \hline
 1 & -1 & -4 & 5 & 3\cr \hline
7 & 9 & 4 & 3 & 0\cr \hline
 8 & 0 &-2 & 0 & E\cr \hline
\end{array},~~~~~~~~~~~~~~~~~~~~~~~~~~~~~~~~~~~~~~~~~~~\begin{array}{|c|c|c|c|c|}\hline
 50  & 52 & 47 & 46 & 47\cr \hline
 53  & 55 & 50 & 56 & 53\cr \hline
 54 & 56 & 51 & 57 & 54\cr \hline
59 & 61 & 56 & 55 & 52\cr \hline
 52 & 52 &47 & 52 & 49\cr \hline
\end{array}. $\\[0.5cm]
{\bf 5. Solving equations system in ${\bf R}_{max}$\\[0.4cm]
{\it Inputs data}\hspace*{9cm}{\it Outputs data}\\[0.2cm]
Number of lines and columns for matrix $A!~~~~~\fbox{4}$\\[0.1cm]
Value for $k!~~~\fbox{10}$\\[0.2cm]
Matrix $~A!~~~~~~~~~~~~~~~~~~~~~$ Matrix $X(0)~~~~~~~~~~~~~~~~~~~~~~$ Matrix $X(10)$\\[0.2cm]
 $\begin{array}{|c|c|c|c|}\hline
 3  & -5 & -9 & 2 \cr \hline
 4  & 8 & 7 & 4 \cr \hline
 -6 & E & 0 & E \cr \hline
1 & 1 & E & 2 \cr \hline
\end{array}~~~~~~~~~~~~~~~~~~ \begin{array}{|c|}\hline
 4  \cr \hline
  3\cr \hline
2\cr \hline
1\cr \hline
\end{array}~~~~~~~~~~~~~~~~~~~~~~~~~~~~~~~~~~~~~\begin{array}{|c|}\hline
 70 \cr \hline
  83  \cr \hline
56   \cr \hline 76\cr \hline
\end{array}.$\\

Author's adresses\\

West University of Timi\c soara,\\
Department of Mathematics, Bd. V. P{\^a}rvan, no. 4, 300223, Timi\c soara, Romania\\
\hspace*{0.7cm} E-mail: mihai31ro@yahoo.com; ivan@math.uvt.ro\\
\end{document}